\documentclass[aps, prb, amsmath, twocolumn, 10pt, superscriptaddress]{revtex4-2}

\usepackage{bbold}
\usepackage{threeparttable}
\usepackage{booktabs}
\usepackage{graphicx}
\usepackage{braket}
\usepackage{upgreek}

\newcommand{\GeSn}{Ge$_{1-x}$Sn$_x$}

\usepackage{xcolor}

\begin{document}

\title{Nuclear spin-free $^{70}$Ge/$^{28}$Si$^{70}$Ge quantum well heterostructures grown on industrial SiGe-buffered wafers}

\author{P. Daoust}
\affiliation{Department of Engineering Physics, \'Ecole Polytechnique de Montr\'eal, Montr\'eal, C.P. 6079, Succ. Centre-Ville, Montr\'eal, Qu\'ebec, Canada H3C 3A7}
\author{N. Rotaru}
\affiliation{Department of Engineering Physics, \'Ecole Polytechnique de Montr\'eal, Montr\'eal, C.P. 6079, Succ. Centre-Ville, Montr\'eal, Qu\'ebec, Canada H3C 3A7}
\author{D. Biswas}
\affiliation{Stewart Blusson Quantum Matter Institute, University of British Columbia, Vancouver, BC, Canada}
\affiliation{Department of Physics and Astronomy, University of British Columbia, Vancouver, BC, Canada}
\author{S. Koelling}
\affiliation{Department of Engineering Physics, \'Ecole Polytechnique de Montr\'eal, Montr\'eal, C.P. 6079, Succ. Centre-Ville, Montr\'eal, Qu\'ebec, Canada H3C 3A7}
\author{E. Rahier}
\affiliation{Department of Engineering Physics, \'Ecole Polytechnique de Montr\'eal, Montr\'eal, C.P. 6079, Succ. Centre-Ville, Montr\'eal, Qu\'ebec, Canada H3C 3A7}
\author{A. Dub\'e-Valade}
\affiliation{Department of Engineering Physics, \'Ecole Polytechnique de Montr\'eal, Montr\'eal, C.P. 6079, Succ. Centre-Ville, Montr\'eal, Qu\'ebec, Canada H3C 3A7}
\author{P. Del Vecchio}
\affiliation{Department of Engineering Physics, \'Ecole Polytechnique de Montr\'eal, Montr\'eal, C.P. 6079, Succ. Centre-Ville, Montr\'eal, Qu\'ebec, Canada H3C 3A7}
\author{M. S. Edwards}
\affiliation{Stewart Blusson Quantum Matter Institute, University of British Columbia, Vancouver, BC, Canada}
\affiliation{Department of Electrical and Computer Engineering, University of British Columbia, Vancouver, BC, Canada}
\author{M. Tanvir}
\affiliation{Stewart Blusson Quantum Matter Institute, University of British Columbia, Vancouver, BC, Canada}
\affiliation{Department of Electrical and Computer Engineering, University of British Columbia, Vancouver, BC, Canada}
\author{E. Sajadi}
\affiliation{Stewart Blusson Quantum Matter Institute, University of British Columbia, Vancouver, BC, Canada}
\affiliation{Department of Electrical and Computer Engineering, University of British Columbia, Vancouver, BC, Canada}
\author{J. Salfi}
\affiliation{Stewart Blusson Quantum Matter Institute, University of British Columbia, Vancouver, BC, Canada}
\affiliation{Department of Electrical and Computer Engineering, University of British Columbia, Vancouver, BC, Canada}
\affiliation{Department of Physics and Astronomy, University of British Columbia, Vancouver, BC, Canada}
\author{O. Moutanabbir}
\email{oussama.moutanabbir@polymtl.ca}
\affiliation{Department of Engineering Physics, \'Ecole Polytechnique de Montr\'eal, Montr\'eal, C.P. 6079, Succ. Centre-Ville, Montr\'eal, Qu\'ebec, Canada H3C 3A7}

\begin{abstract}

 The coherence of hole spin qubits in germanium planar heterostructures is limited by the hyperfine coupling to the nuclear spin bath due to $^{29}$Si and $^{73}$Ge isotopes. Thus, removing these nuclear spin-full isotopes is essential to extend the hyperfine-limited coherence times needed to implement robust quantum processors. This work demonstrates the epitaxial growth of device-grade nuclear spin-free $^{70}$Ge/$^{28}$Si$^{70}$Ge heterostructures on industrial SiGe buffers while minimizing the amounts of highly purified $^{70}$GeH$_4$ and $^{28}$SiH$_4$ used. The obtained $^{70}$Ge/$^{28}$Si$^{70}$Ge heterostructures exhibit a dislocation density of $5.3 \times 10^{6}~\mathrm{cm}^{-2}$ and an isotopic purity exceeding $99.99\%$, with carbon and oxygen impurities below the detection sensitivity, as revealed by atom probe tomography. Magneto-transport measurements on gated Hall bars demonstrate effective gate control of hole density in nuclear spin-free quantum wells. Negative threshold gate voltages confirm the absence of intentional doping in the wells, while Hall and Shubnikov--de Haas analyses yield consistent carrier densities ($\sim 1.4 \times 10^{11}~\mathrm{cm}^{-2}$) and high mobilities ($\sim 2.4 \times 10^{5}~\mathrm{cm}^{2}/\mathrm{Vs}$). Mobility trends reveal interface-trap-limited scattering and percolation concentration below $7 \times 10^{10}~\mathrm{cm}^{-2}$. These analyses, along with atomic-level studies, confirm the high quality of epitaxial $^{70}$Ge/$^{28}$Si$^{70}$Ge heterostructures and their relevance as a platform for long-coherence spin qubits.
\end{abstract}

\maketitle

\footnotetext[1]{Details on the $k\cdot p$ model, the perturbative framework, the cubic Rashba parameters, the consistency of the quantum channel theory at large $l_x$ and the \GeSn{} material parametrization are provided in the Supplemental Material}

\paragraph*{}
Germanium (Ge) and its alloys have emerged as a promising platform to implement scalable quantum processors, leveraging robust Ge-based hole spin qubits~\cite{Scappucci2021,Vecchio2025,Borsoi2024,Hsiao2024,Ivlev2024,Vecchio2023,Lawrie2023,vanRiggelen2021,Hendrickx2021,Hendrickx2020N,Rotaru2025,Hardy2019,Jirovec2021,Vecchio2024,Kaul2025,Bosco2021,bogan2019,Wang2016,Watzinger2018,Wang2022}. Besides its compatibility with silicon processing standards, Ge offers a quiet quantum environment to engineer hole spin states within scalable device structures with favorable properties for qubit control and coherence. The large spin–orbit interaction (SOI) in Ge enables all-electrical manipulation of spin states, i.e., control of spin via the charge degree of freedom, eliminating the need for oscillating magnetic fields and allowing faster and more scalable qubit control~\cite{Rashba1991}. The pronounced SOI provides tunable $g$-factors through electric fields, offering versatile pathways to optimize spin coherence and coupling in quantum dot architectures~\cite{Scappucci2021}. Moreover, it is generally expected that holes experience a weaker hyperfine interaction than electrons owing to the predominantly $p$-type symmetry of their wavefunctions, which vanishes at the nucleus. This property initially provided an additional motivation to pursue the development of hole-spin qubits. However, theoretical studies have shown that the hyperfine coupling for holes can be only an order of magnitude smaller than that of electrons~\cite{Testelin2009,Philippopoulos2020}, and in some cases comparable to that in silicon ~\cite{Philippopoulos2020}. Furthermore, the $p$-orbital character combined with $d$-orbital hybridization gives rise to pronounced anisotropy in the hole hyperfine interaction, a feature absent in electron spins~\cite{Testelin2009}.

The extent of the hyperfine interactions was recently addressed in a study probing both charge noise and hyperfine-mediated magnetic noise in hole spin qubits embedded in planar strained Ge/Si$_x$Ge$_{1-x}$ heterostructures~\cite{Stehouwer2025}. That work, based on echo envelope modulations and noise-spectrum fitting, revealed that residual coupling to $^{73}$Ge and $^{29}$Si nuclear spin bath contributes significantly to dephasing, setting an approximate $T_2^*$ bound of $\sim 1~\mu$s under their natural isotopic composition. The study further estimates that fully isotopically purified Ge and surrounding Si$_{1-x}$Ge$_x$ barriers could push the hyperfine-limited coherence times into the tens to hundreds of microseconds regime. Furthermore, a recent experimental study demonstrated that, due to the anisotropy of the hyperfine coupling and the $g$-tensor, in the presence of inhomogeneous strain, out-of-plane magnetic fields are optimal for mitigating decoherence induced by electric field fluctuations, whereas in-plane magnetic fields are more effective in suppressing decoherence arising from the nuclear spin bath~\cite{Hendrickx2024}. These studies show that removing nuclear spins is a critical pathway to improve coherence for hole spin qubits. These observations confirm that developing nuclear spin-free Ge heterostructures is essential toward practical hole spin qubits.

\begin{figure*}[t]
     \centering
  \includegraphics[width=0.66\textwidth]{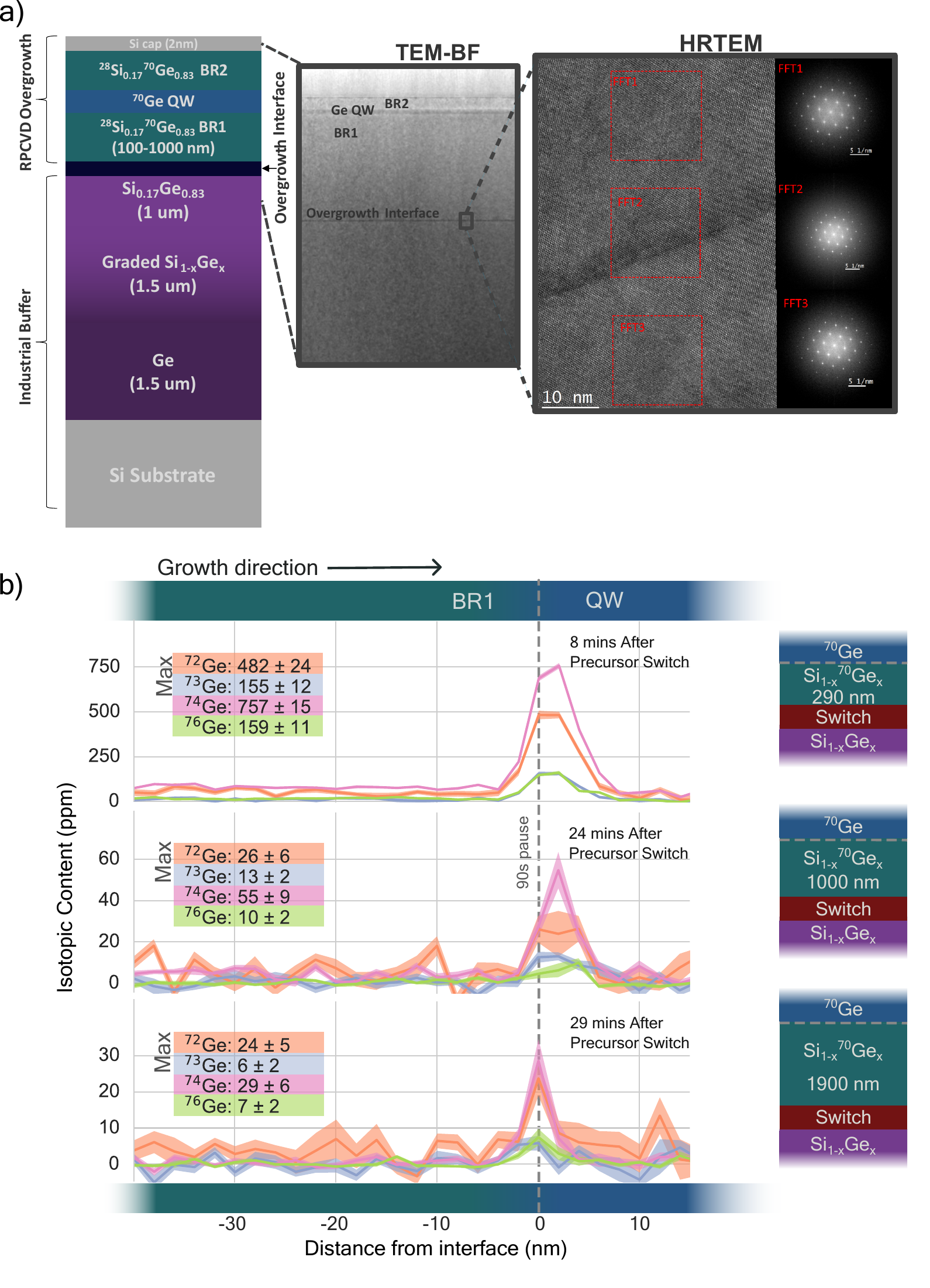}
\caption{(a) Schematic illustration of the growth stack used to grow Ge/SiGe heterostructures on industrial SiGe buffers. Inset~1: Low-magnification TEM image of a typical $^{70}$Ge/$^{28}$Si$^{70}$Ge heterostructure. Inset~2: High-resolution TEM micrograph recorded at the regrowth interface. (b) Atom probe tomography (APT) profiles of Ge isotopes in Ge/SiGe heterostructures obtained under different growth protocols.}
  \label{fig:wide}
\end{figure*}

Recent attempts to grow isotopically controlled $^{70}$Ge/$^{28}$Si$^{70}$Ge heterostructures~\cite{Moutanabbir2024}, using purified $^{70}$GeH$_4$ and $^{28}$SiH$_4$ precursors, demonstrate that cross-contamination from natural precursors in the growth reactor makes the complete removal of nuclear spin-full nuclei $^{29}$Si and $^{73}$Ge rather difficult and somewhat impractical. The growth of planar Ge/SiGe heterostructures on silicon typically involves the initial growth of a thick strained-relaxed Si$_{1-x}$Ge$_x$ buffer layer, ideally using conventional precursors with natural isotopic content (e.g., $^{Nat}$GeH$_4$ and $^{Nat}$SiH$_4$). The heterostructure, consisting of barriers and a quantum well (QW), is subsequently grown by switching to isotopically purified precursors. However, traces of natural precursors in the reactor can hardly be eliminated during this last step, leading to the undesirable incorporation of nuclear spin-full species in the grown $^{70}$Ge/$^{28}$Si$^{70}$Ge heterostructures~\cite{Moutanabbir2024}. Although this reservoir effect can be greatly suppressed by using exclusively $^{70}$GeH$_4$ and $^{28}$SiH$_4$ to grow the entire several-micrometer-thick stack (e.g., Figure~1(a)), including the relaxed Si$_{1-x}$Ge$_x$ buffers, this approach remains infeasible because it requires excessive amounts of rare and costly isotopically enriched precursors.

Herein, we circumvent these challenges by establishing protocols for the epitaxial growth of nuclear spin-free $^{70}$Ge/$^{28}$Si$^{70}$Ge heterostructures directly on 200~mm industrial Si$_x$Ge$_{1-x}$-buffered silicon wafers, as illustrated in Figure~1(a). A meticulous surface-cleaning process was developed to properly condition the Si$_x$Ge$_{1-x}$ surface to enable the growth of $^{70}$Ge/$^{28}$Si$^{70}$Ge heterostructures employing a mix of wet chemical treatment and \textit{in situ} annealing, which we subsequently describe in detail. Note that the cleaning processes previously developed for Ge \cite{Ponath2017} and Si$_x$Ge$_{1-x}$  \cite{Shimura2024} surfaces are not effective for Ge-rich Si$_x$Ge$_{1-x}$ surfaces investigated here. The quality of the heterostructures was assessed through detailed investigations of their microstructure and atomic-level three-dimensional isotopic composition using transmission electron microscopy (TEM), X-ray diffraction (XRD), atomic force microscopy (AFM), and atom probe tomography (APT). Magneto-transport measurements further confirm the high crystallinity and excellent interfacial quality of the obtained nuclear spin-free $^{70}$Ge/$^{28}$Si$^{70}$Ge heterostructures.

The growth was carried out on the pre-conditioned Si$_x$Ge$_{1-x}$ surfaces in a reduced-pressure chemical vapor deposition reactor equipped with isotopically purified $^{70}$GeH$_4$ (isotopic purity $>99.9\%$) and $^{28}$SiH$_4$ (isotopic purity $>99.99\%$) precursors. These precursors contain only traces of other Si ($^{29}$Si and $^{30}$Si) and Ge ($^{72}$Ge, $^{73}$Ge, $^{74}$Ge, and $^{76}$Ge) isotopes, with a combined total content below $0.006~\mathrm{at.\%}$ in each precursor. The presence of chemical contaminants, including residual hydrides, is also negligible, with an average content of less than $0.06~\mu\mathrm{mol/mol}$. After systematic studies (see Methods), $550~^{\circ}\mathrm{C}$ is identified as the optimal growth temperature that yields high-quality growth and sharp interfaces. To assess the ability of this process to eliminate cross-contamination, reference samples were also investigated. In these samples, both the heterostructure and the buffer layer were grown in the same reactor by switching from natural to purified precursors~\cite{Moutanabbir2024}.

\begin{figure*}[t]
     \centering
  \includegraphics[width=\textwidth]{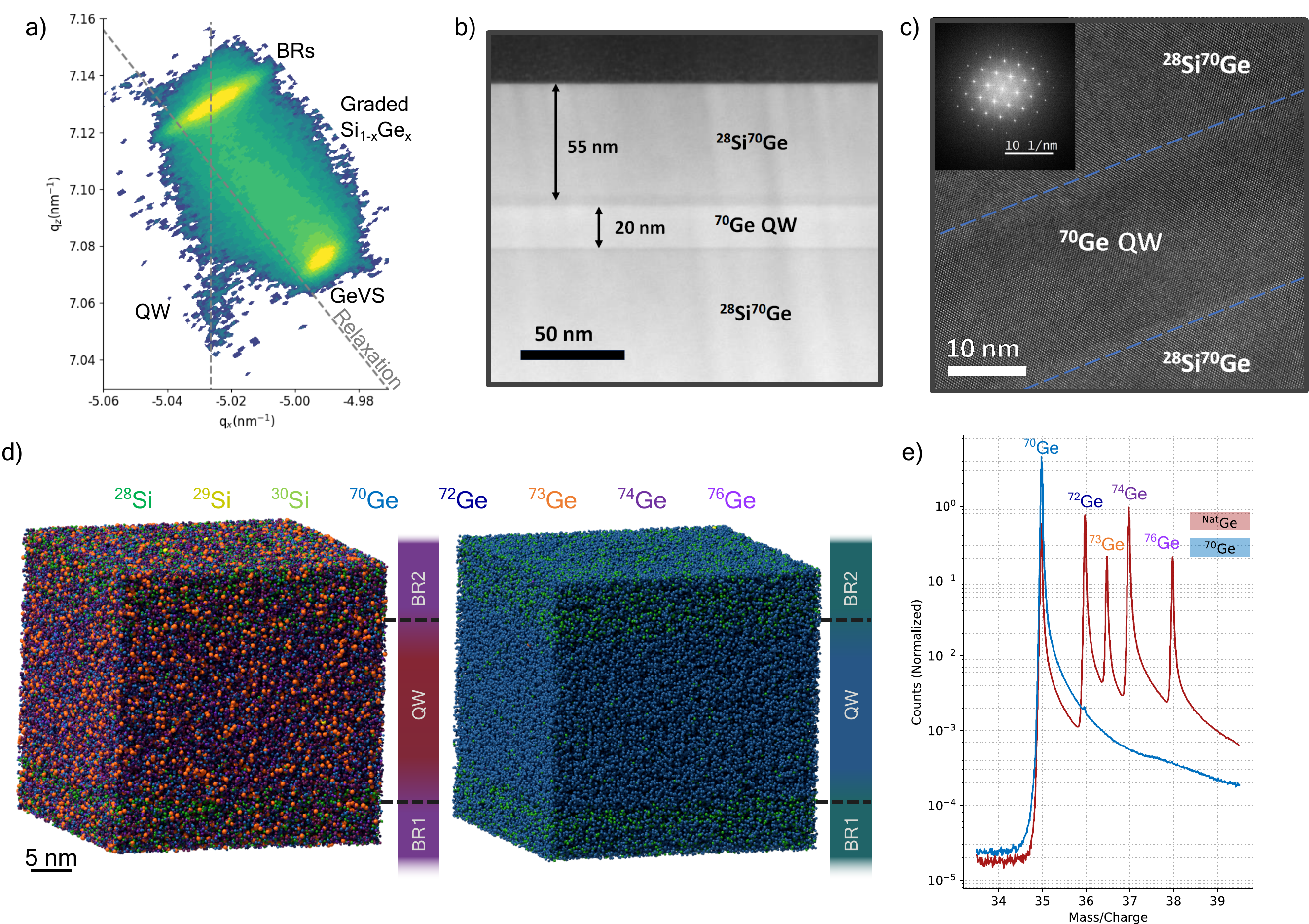}
\caption{(a) XRD-RSM recorded around the ($\overline{2}\,\overline{2}\,4$) reflection for an optimized $^{70}$Ge/$^{28}$Si$^{70}$Ge heterostructure showing the relaxation line (diagonal dashed line) and quantum well pseudomorphism line (vertical dashed line). (b) Annular dark-field TEM image of an as-grown $^{70}$Ge/$^{28}$Si$^{70}$Ge heterostructure with a 20~nm-thick quantum well and a 55~nm-thick top barrier. (c) High-resolution TEM image of the $^{70}$Ge quantum well and its interfaces with the SiGe barriers. (d) Three-dimensional isotope-by-isotope maps of QW structures grown with natural precursors (left) and isotopically purified precursors (right). (e) APT mass spectra showing the detected Ge isotopes for the heterostructures in (d).}
  \label{fig:wide}
\end{figure*}

Figure~1(b) exhibits the APT profiles of the isotopic content along the growth direction from the Si$_{1-x}$Ge$_x$ buffer layer up to the $^{70}$Ge quantum well (QW) for the reference samples for which the switch to purified precursor took place at different depths from the QW interface (i.e., different times from the QW growth). For simplicity, only the profiles of Ge isotopes are displayed. Note that a 90~s growth pause right before the QW growth was introduced to ensure a sharp Si$_x$Ge$_{1-x}$/Ge interface. The APT profiles clearly reveal the incorporation of undesired Ge isotopes within the QW. Even after growing a 1.9~$\mu$m-thick Si$_{1-x}$Ge$_x$ layer using purified precursors, corresponding to 29~min after halting the supply of natural precursors, a measurable isotopic carry-over remains (bottom profile). As expected, this reservoir effect becomes more pronounced when the growth time after precursor switching is shorter. For example, only 8~min after the switch (top profile), the peak concentration of $^{73}$Ge in the QW exceeds 150~ppm.

Eliminating this cross-contamination can be achieved by establishing the growth of $^{70}$Ge/$^{28}$Si$^{70}$Ge heterostructures on Si$_x$Ge$_{1-x}$ wafers (Figure~1(a)). Surface conditioning prior to growth is the main challenge to grow high-quality heterostructures on \textit{ex situ} grown Si$_x$Ge$_{1-x}$ buffers. In general, cleaning Si$_x$Ge$_{1-x}$ surfaces prior to epitaxy can be a complicated and daunting undertaking due to the complex nature of Si and Ge native oxides and related surface chemistry \cite{Shimura2024,Ponath2017}. The mixed SiO$_2$–GeO$_2$ oxide exhibits non-uniform thermal desorption behavior: while GeO$_2$ is volatile and decomposes around 400–600~$^{\circ}$C, SiO$_2$ remains stable at higher temperatures. Moreover, GeO$_2$ can react with Si to form SiO$_2$ and elemental Ge, leading to Ge enrichment and surface roughening during annealing. Ge also segregates to the surface under thermal or chemical treatment, promoting reoxidation and non-uniform hydrogen termination after HF-based cleans. These effects, compounded by a narrow thermal window for oxide desorption, make achieving an atomically clean, ordered Si$_x$Ge$_{1-x}$ surface difficult compared to pure Si or Ge. 

Here, a multi-step cleaning process involving diluted HF and HCl cleans, as well as \textit{in situ} hydrogen anneals at high temperature, is performed before the epitaxial growth of the $^{70}$Ge/$^{28}$Si$^{70}$Ge heterostructure. The optimized surface preparation protocol is summarized in the Methods section and in Table~S1 of the Supplementary Information (SI).

The epitaxial growth on Si$_x$Ge$_{1-x}$ buffers starts with an initial boost of $^{28}$SiH$_4$ to prevent the formation of Ge three-dimensional islands. Once the growth of the first barrier (BR1) is completed, the precursor flows were cut for approximately 30~s, during which the reactor is purged with 2700~sccm of hydrogen flow. The growth of the $^{70}$Ge quantum well (QW) then follows with a subsequent 30~s hydrogen purge of the reactor. To ensure a sharp interface between the well and the second barrier (BR2), the $^{28}$SiH$_4$ precursor flow is momentarily increased through a second boost step. Finally, a $\sim 2$~nm capping layer of $^{28}$Si is grown to protect the purified heterostructure with a stable native oxide. The heterostructure dimensions were optimized following systematic growth experiments to obtain a uniform thickness for each layer within a range ensuring heavy-hole confinement while keeping its wavefunction away from the noisy surface, as confirmed by $\mathbf{k}\cdot\mathbf{p}$ theory calculations (not shown).

Figure~1(a) (inset) displays a low-magnification TEM image of a representative heterostructure, confirming that the growth protocol described above yields heterostructures free of extended defects (at the TEM scale) with a uniform thickness. Note that the homoepitaxial layer grown directly on the strain-relaxed Si$_x$Ge$_{1-x}$ buffer is also defect-free. Indeed, high-resolution TEM imaging and related diffraction patterns at the regrowth interface demonstrate that the grown layers display the same lattice structure as that of the industrial buffers, as exemplified in Fig.~1(a) (inset). The noticeable change in contrast at the interface is attributed to the formation of a Ge-rich layer at the onset of growth, as confirmed by APT (SI, Sec.~4).

Further insights into the basic structural properties of the grown heterostructures are obtained from high-resolution XRD reciprocal space mapping (XRD-RSM) analyses. Figure~2(a) displays a typical XRD-RSM recorded around the ($\overline{2}\,\overline{2}\,4$) reflection for an optimized heterostructure with a 20~nm-thick Ge QW and a 55~nm-thick Si$_x$Ge$_{1-x}$ BR2. Figures~2(b) and 2(c) show annular dark-field TEM and high-resolution TEM images of this heterostructure, respectively. Although the quantum well is only 20~nm in thickness, its signature can be clearly observed in the XRD-RSM under the sharp overlapping reflections of the Si$_x$Ge$_{1-x}$ ($x=0.17$) buffer and barriers. As there is no deviation from the pseudomorphism line (vertical dashed line), the grown QW did not relax and thus remains fully compressively strained. The composition of the barriers was determined using a Vegard-type model~\cite{Xu2017}, yielding $^{28}$Si$_{0.17}$Ge$_{0.83}$, identical to that of the underlying relaxed buffers.

The roughness of as-grown heterostructures was investigated using AFM (SI). Prior to growth, the industrial buffers exhibit an average and a root-mean-square roughness of $R_\mathrm{a}=1.5$~nm and $R_\mathrm{q}=2.0$~nm, respectively. Surface conditioning slightly increases these values to $R_\mathrm{a}=2.1$~nm and $R_\mathrm{q}=2.7$~nm, which remain unchanged after the $^{70}$Ge/$^{28}$Si$^{70}$Ge heterostructure growth. This behavior is consistently observed regardless of the thickness of the first barrier (BR1), which was varied between 100~nm and 1~$\mu$m. Standard Secco etch~\cite{Secco1972} defect delineation studies were subsequently performed to probe the dislocation density in the grown materials. Etch pits, attributed to threading dislocations, were found to show densities of $7.8\times10^{6}~\mathrm{cm}^{-2}$ and $5.3\times10^{6}~\mathrm{cm}^{-2}$ for the industrial buffers and grown heterostructures, respectively. These dislocation densities are consistent with the literature~\cite{Abbadie2009} and confirm the high quality of the grown heterostructures. More details on these studies are provided in Sec.~S3 of the SI.

 \begin{figure*}[t]
     \centering
  \includegraphics[width=\textwidth]{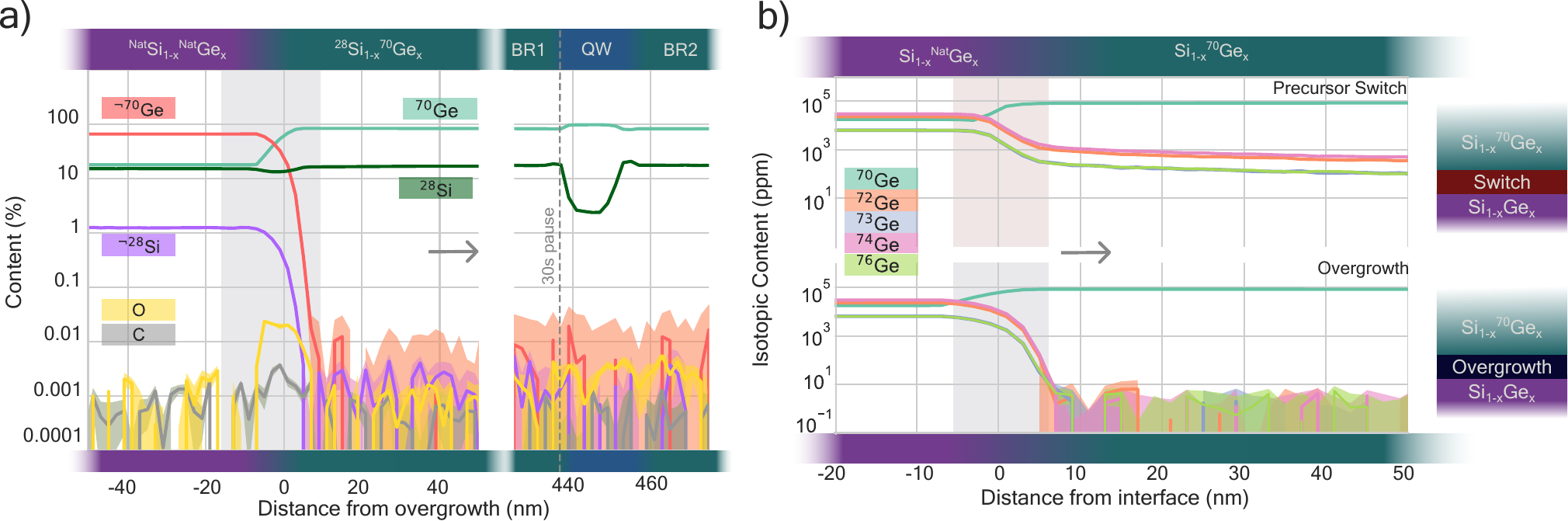}
\caption{(a) APT compositional profiles near the overgrowth interface showing carbon (C) and oxygen (O) impurities as well as Si and Ge isotopes other than $^{28}$Si and $^{70}$Ge, labeled as $^{\neg28}$Si and $^{\neg70}$Ge. (b) APT profiles of Ge isotopes in heterostructures grown on SiGe-buffered silicon (bottom) and in a heterostructure grown by switching from natural to purified precursors (top). Note that the content axis is plotted on a logarithmic scale. The arrows indicate the growth direction.}
  \label{fig:wide}
\end{figure*}

 To elucidate the isotopic and chemical purity of the as-grown heterostructures, laser-assisted atom probe tomography (APT) measurements were carried out using a picosecond laser with a wavelength of 257~nm and a pulse energy of 25--45~pJ. The analyses were performed at a base temperature of 25~K. Figure~2(d) displays representative APT three-dimensional isotope-by-isotope reconstructed maps of a typical natural Ge/SiGe heterostructure (left) and an optimized isotopically controlled $^{70}$Ge/$^{28}$Si$^{70}$Ge heterostructure (right). In the natural heterostructure, nuclear spin-full isotopes $^{29}$Si and $^{73}$Ge are clearly visible in both the quantum well (QW) and the barrier layers. The average distance between nuclear spins in this heterostructure is 0.3--0.4~nm. In contrast, $^{29}$Si and $^{73}$Ge are not detected in the map of the $^{70}$Ge/$^{28}$Si$^{70}$Ge heterostructure. The Ge mass-to-charge spectra corresponding to these maps are shown in Figure~2(e). Note that in the natural Ge/SiGe heterostructure, five peaks associated with naturally occurring Ge isotopes are observed, whereas only a single peak is detected for $^{70}$Ge/$^{28}$Si$^{70}$Ge, confirming the isotopic purity of these heterostructures.

Figure~3(a) displays the composition profiles of $^{70}$Ge and $^{28}$Si along the growth direction from about 100~nm underneath the growth interface up to the surface. The profiles of C and O impurities are also displayed. The $^{70}$Ge and $^{28}$Si contents in the heterostructure are 83\% and 17\%, respectively, and remain stable throughout the barrier thickness. These values are consistent with the XRD-RSM data described above. Furthermore, the composition of the alloy stabilizes within about 30~nm beyond the interface, suggesting that the thickness of BR1 could be further reduced below 100~nm. Apart from a peak of O with a content below 300~ppm at the growth interface, attributed to residual oxide from surface preparation, the levels of O and C impurities in the heterostructure are too close to the APT sensitivity limit (a few ppm) to be meaningful. The concentrations of isotopes other than $^{70}$Ge and $^{28}$Si, labeled as $^{\neg28}$Si and $^{\neg70}$Ge, drop rapidly within a few nanometers of the regrowth interface to reach $^{\neg28}$Si = 8~ppm and $^{\neg70}$Ge = 40~ppm  on average over the first 200~nm beyond the interface. This corresponds to an isotopic purity, defined as the ratio of the content of an isotope to the total element content, higher than $\ 99.99\%$. This rules out any diffusion of material from the buffers or cross-contamination from past growth experiments. This is further evidenced in Figure~3(b), which compares the profiles of Ge isotopes for a heterostructure overgrown on a SiGe-buffered wafer (bottom) to that of a heterostructure grown by switching the precursors from natural to isotopically enriched sources (top). In particular, in the former, there is no tail associated with the reservoir effect or residues at the growth front, as observed when the buffers are grown in the same reactor using natural precursors.

The analyses above demonstrate that the introduced growth protocol yields $^{70}$Ge/$^{28}$Si$^{70}$Ge heterostructures of high crystalline quality and high chemical and isotopic purities, thus laying the foundation for transport studies. To this end, gated six-terminal Hall bars (HBs) were fabricated using optical lithography on the purified heterostructures. Figures~4(a) and 4(b) illustrate the device layout and its cross-section, respectively. The devices investigated here have a length $L = 300~\mu\mathrm{m}$ between the voltage contacts and a width $W = 100~\mu\mathrm{m}$. The HBs were measured in a cryo-free variable temperature insert (VTI) at a temperature of approximately 1.4~K, in the bore of an 8~T superconducting solenoid. 

\begin{figure*}[t]
     \centering
  \includegraphics[width=0.8\textwidth]{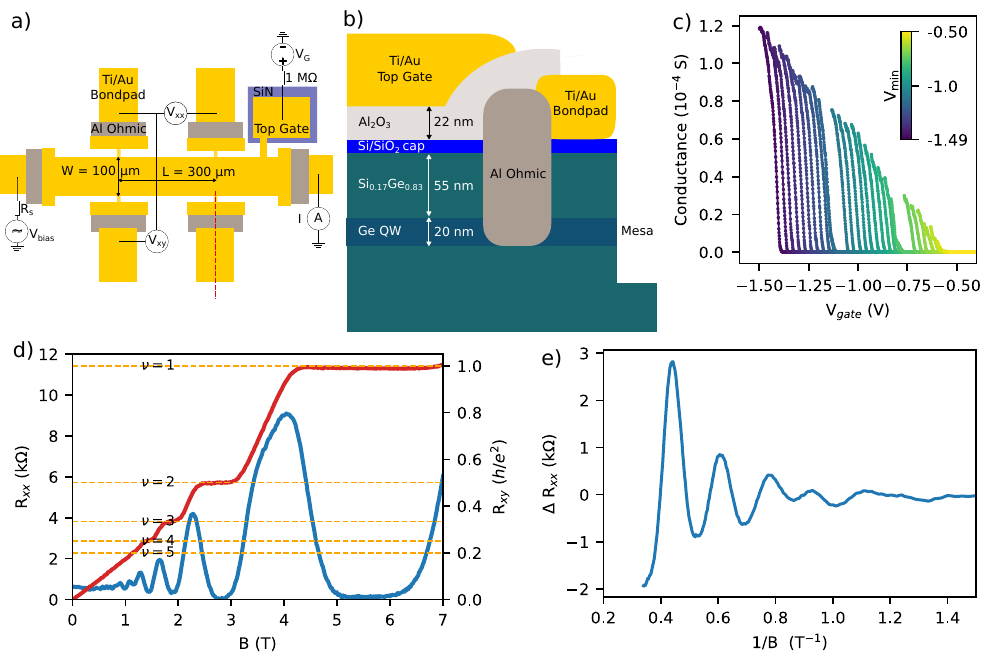}
\caption{(a) Measurement setup and schematic of a gated Hall bar (HB), showing the excitation voltage $V_{\mathrm{bias}}$ and measured quantities $V_{xx}$ and $V_{xy}$, with $R_s = 10~\mathrm{M}\Omega$ and $V_{\mathrm{bias}} = 40~\mathrm{mV}$ for magnetotransport measurements, and measured current $I$ with $R_s = 0~\Omega$ and $V_{\mathrm{bias}} = 50~\mu\mathrm{V}$ for two-point measurements. The red dashed line indicates the cross-sectional area through the device shown in (b). (b) Cross-section of the Ge/SiGe heterostructure and gated Hall bar. (c) Two-point measurement showing turn-on conductance curves at 1.38~K for a representative HB device, recorded for a range of $V_{\mathrm{min}}$ sweep values from $-0.5$~V (yellow) to $-1.5$~V (blue). (d) Magnetotransport up to $B_{\perp} = 7$~T at 1.38~K and $V_{\mathrm{gate}} = -2.55$~V, showing $R_{xy}$ (red curve) and $R_{xx}$ (blue curve). The orange dashed lines indicate the integer Landau level filling factors. (e) The corresponding Shubnikov–de~Haas (SdH) oscillations are extracted by subtracting an even polynomial background from the data in (d).
}
  \label{fig:wide}
\end{figure*}

For magneto-transport measurements, we measured the longitudinal $R_{xx} = V_{xx}/I$ and transversal $R_{xy} = V_{xy}/I$ resistances by applying an oscillating bias voltage $V_{\mathrm{bias}} = 40~\mathrm{mV}$ to a resistor $R_s = 4~\mathrm{M}\Omega \gg R_{xx}$ in series with the device, resulting in a current $I = V_{\mathrm{bias}}/R_s = 10~\mathrm{nA}$. Gate voltage $V_G$ and temperature were kept constant during sweeps of the perpendicular magnetic field $B_{\perp}$. The gate voltage $V_G$ was swept from 0 to $V_{\mathrm{min}}$, then from $V_{\mathrm{min}}$ back to zero, for values of $V_{\mathrm{min}}$ starting at zero and becoming increasingly negative. This procedure was performed to study the effect of gate-induced charging of interface traps, which shift the threshold voltage. Measurements were repeated by releasing these trapped charges and restoring the unshifted turn-on voltage by cycling the VTI temperature to 300~K and back to 1.4~K.

For the HB measurements, we first verified the ability of a voltage applied to the gate of the HB to control the conductance and hole concentration in the quantum well (QW). Measurements are shown in Fig.~4(c) for a representative gated HB, as a function of $V_G$, for several values of $V_{\mathrm{min}}$. When $V_G$ is swept from 0 to $V_{\mathrm{min}}<0$, and back to 0, all devices in the fabrication run turned on at a gate voltage of approximately $-0.6$~V. Forward and backward sweeps for specific $V_{\mathrm{min}}$ values exhibited a hysteresis, confirming the presence of charge trapping, and as $V_{\mathrm{min}}$ is decreased, the turn-on $V_G$ shifts to more negative values, confirming an overall increase in positively trapped interface charges. This has previously been reported in natural Ge/Si$_x$Ge$_{1-x}$ heterostructures~\cite{Massai2023}. The threshold voltage for turn-on being negative verifies that the isotopically purified QWs are free of carriers at zero gate voltage, i.e., there is no unintentional doping.

For simplicity, we first focus on magneto-transport measurements at a fixed gate voltage $V_G$ and carrier concentration $p$. The data shown in Figure~4(d) exhibit, at very low magnetic fields, the Hall effect, i.e., the transverse resistance $R_{xy}$ is proportional to $B_{\perp}$. From the expression for the Hall resistance $R_{xy} = V_{xy}/I_{xx} = B_{\perp}/(ep)$, where $e$ is the elementary charge, we extract a hole concentration $p = 1.49\times10^{11}~\mathrm{cm}^{-2}$. Increasing the magnetic field $B_{\perp}$ further, we observe Shubnikov–de~Haas (SdH) oscillations, periodic in $1/B_{\perp}$, as Landau levels pass through the Fermi energy and modulate the longitudinal resistance. For magnetic fields around 2.7~T and higher, when Landau levels pass through the Fermi energy, we observe the integer quantum Hall effect, with well-defined transverse resistance $R_{xy}$ plateaus and zero longitudinal resistance $R_{xx}$ for the $\nu=2$ and $\nu=1$ Landau levels at $p = 1.49\times10^{11}~\mathrm{cm}^{-2}$ and $T = 1.38$~K. The SdH oscillations $\Delta R_{xx}$ are plotted versus $1/B_{\perp}$ in Figure~4(e) at the same $V_G = -2.55$~V. Assuming a circular Fermi surface, the lowest-harmonic SdH oscillation observed is periodic in $1/B$ with period $2\pi p h / e$ ~\cite{Kuppersbusch2017}, and using this relation, we find a hole concentration $p = 1.41\times10^{11}~\mathrm{cm}^{-2}$, which is very close to that extracted from the Hall effect, as expected for a QW with negligible bulk transport contribution.

Next, we discuss the electronic hole mobility $\mu_p$ and concentration $p$, measured as a function of gate voltage to assess the quality of the accumulation-mode quantum well HBs. The gate voltage $V_G$ was swept from 0 to $V_{\mathrm{min}}$, and back to 0, for different $V_{\mathrm{min}}$ values. At each $V_G$, the magnetic field $B_{\perp}$ was varied to obtain $R_{xx}$ and $R_{xy}$, and the Hall mobility $\mu_p$ and hole concentration $p$ were obtained from the relations $R_{xx} = (W/L)/(\mu_p p e)$ and $R_{xy} = B_{\perp}/(ep)$. Data for $p$ and $\mu_p$ are shown in Figure~5(a) for $V_G$ ranging from $-2.14$~V to $-2.2$~V at $V_{\mathrm{min}} = -2.2$~V. The carrier concentration $p$ increases approximately linearly as $V_G$ decreases, as expected. Starting from lower hole concentration, the mobility initially increases with increasing $p$, reaches a maximum value of about $2.2\times10^{5}~\mathrm{cm}^{2}/(\mathrm{V\,s})$, and then decreases for further increases in $p$. The overall upward trend at low concentration is similar to what is reported in natural Ge/Si$_x$Ge$_{1-x}$ heterostructures~\cite{Massai2023}, and can be attributed to better screening of Coulomb impurities as carrier concentration increases. Further decrease in gate voltage (i.e., increased carrier concentration) populates positive charge traps at, e.g., the Si$_x$Ge$_{1-x}$/Si/SiO$_2$ interface, thereby reducing carrier mobility.

Reliable Hall voltages were difficult to extract for gate voltages just below the turn-on voltage, where the Hall bar resistance $R_{xx}$ varied on a timescale much longer than that required to sweep the magnetic field. We attribute this slow variation of device resistance to slow changes in the population of trapped charges (and therefore hole concentration $p$) at the semiconductor–dielectric interface, Si$_x$Ge$_{1-x}$/Si/SiO$_2$/Al$_2$O$_3$. The change in hole concentration $p$ introduces a slow drift in $R_{xx}$ and $R_{xy} = B_{\perp}/(ep)$, making the extraction of $p$ and $\mu_p$ as a function of $B_{\perp}$ unreliable.

\begin{figure}[t]
  \centering
  \includegraphics[width=\linewidth]{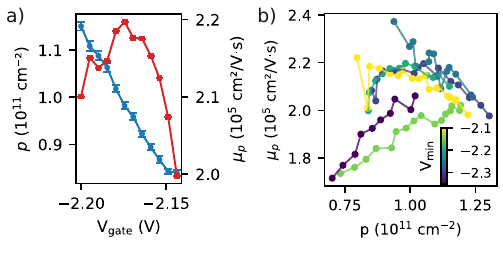}
  \caption{(a) Extracted carrier concentration $p$ (blue curve) and mobility $\mu_p$ (red curve) for $V_{\mathrm{min}} = -2.2$~V, as a function of $V_G$. (b) Extracted Hall mobility versus extracted Hall carrier concentration for different values of $V_{\mathrm{min}}$ ranging from $-2.1$~V (yellow) to $-2.35$~V (blue).
}
\end{figure}

To study the impact of interface charges on the transport properties, we plot the dependence of the extracted Hall mobility $\mu_p$ on the extracted carrier concentration $p$, for different values of $V_{\mathrm{min}}$, during the negative sweep of $V_G$ from 0 to $V_{\mathrm{min}}$. Recall that for smaller values of $V_{\mathrm{min}}$, the turn-on voltage shifts negatively, confirming that trapped charges become more positive. The variation of turn-on voltage with $V_{\mathrm{min}}$ is plotted in Figure~S1(a). The Hall mobility as a function of carrier concentration for different values of $V_{\mathrm{min}}$ is shown in Figure.~5(b). All data show that the mobility increases with decreasing carrier concentration, starting from $p \approx 1.2\times10^{11}~\mathrm{cm}^{-2}$, and for most curves, we observe saturation of the mobility and a downturn as the concentration becomes lower. The exact values of the mobility at low hole concentration do not follow a systematic trend, possibly due to variations in the spatial distribution of trapped charges or a failure to reach a steady-state trapping distribution at low carrier concentrations.

The percolation concentration $p_p$ is an important figure of merit for laterally gated quantum dots formed in quantum wells (QWs). This is because the percolation concentration quantifies the disorder at low carrier densities relevant to, e.g., few-hole quantum dots. We measured monotonically increasing conductance for carrier concentrations as low as $7\times10^{10}~\mathrm{cm}^{-2}$, and therefore infer that the percolation concentration lies below this value. A reliable extrapolation of the conductance versus concentration data to zero conductance is, however, difficult with our data (see Figure~S1(b)).

We now briefly comment on how the properties of the purified material compare to literature values. The peak mobility of $2.4\times10^{5}~\mathrm{cm}^{2}/(\mathrm{V\,s})$ at a carrier concentration of $9\times10^{10}~\mathrm{cm}^{-2}$ and a percolation concentration below $7\times10^{10}~\mathrm{cm}^{-2}$ indicate high material quality suitable for studying single-hole coherence in laterally gated quantum dots, in these nuclear spin-free $^{70}$Ge/$^{28}$Si$^{70}$Ge heterostructures. Reported values of mobility (percolation concentration) in the literature range from $7\times10^{4}$ to $5\times10^{5}~\mathrm{cm}^{2}/(\mathrm{V\,s})$ ($4\times10^{10}$ to $1.2\times10^{11}~\mathrm{cm}^{-2}$) in undoped QWs with barriers of comparable composition at temperatures around 1~K ~\cite{Sammak2019,Massai2024,Li2023,Zhang2024,Sangwan2024,kong2023}. Representative literature values are summarized in Table~S2 of the Supplementary Information (SI). Generally, higher mobilities (lower percolation concentrations) are expected for larger QW depths, as this brings the QW further from the Si$_x$Ge$_{1-x}$/Si/SiO$_2$/Al$_2$O$_3$ interface. 

It is worth noting that the above comparisons are limited due to small differences in the heterostructure layer thicknesses, barrier compositions, and related strain. Nevertheless, some additional observations are instructive. A trap population independent of gate voltage (carrier concentration) would likely yield a mobility that increases with carrier concentration due to improved screening of defects. However, this stands in contrast to our observations, as the mobility consistently plateaus and then decreases with increasing carrier concentration, regardless of $V_{\mathrm{min}}$. We therefore speculate that filling of interface traps via the gate, which increases as $V_G$ decreases (carrier concentration increases), produces a scattering potential that limits the mean free path at high carrier concentrations. It is thus possible, at least for the highest hole concentrations reported, that the parameters measured here for HBs are limited by traps at an interface formed during device processing, rather than by the intrinsic heterostructure interfaces.

In summary, to prevent cross-contamination from natural precursors and minimize the use of isotopically enriched gases without compromising crystalline quality, this work demonstrates the epitaxial growth of nuclear spin-free $^{70}$Ge/$^{28}$Si$^{70}$Ge heterostructures on industrial SiGe buffers using reduced-pressure chemical vapor deposition with highly purified $^{70}$GeH$_4$ ($>99.9\%$) and $^{28}$SiH$_4$ ($>99.99\%$). The resulting heterostructures exhibit a dislocation density of $5.3\times10^{6}~\mathrm{cm}^{-2}$ and isotopic purity above $99.99\%$, free of carbon and oxygen impurities, as confirmed by atom probe tomography. Hall bar measurements show precise gate control of hole density and high mobilities ($\sim2.4\times10^{5}~\mathrm{cm}^{2}/\mathrm{V\,s}$), with transport limited by interface traps and percolation below $7\times10^{10}~\mathrm{cm}^{-2}$. These results establish high-purity $^{70}$Ge/$^{28}$Si$^{70}$Ge heterostructures as a promising platform for long-coherence hole spin qubits, free from hyperfine interactions with $^{29}$Si and $^{73}$Ge nuclei.

\vspace{2em}

\section*{\textbf{ACKNOWLEDGMENTS}}

The authors thank J.~Bouchard for technical support with the RPCVD system. O.M. acknowledges support from the Natural Sciences and Engineering Research Council of Canada (NSERC; Discovery Grants, Alliance International Quantum, and CQS2Q Consortium), the Canada Research Chairs Program, the Canada Foundation for Innovation (CFI), Mitacs, PRIMA Québec, and Defence Canada (Innovation for Defence Excellence and Security, IDEaS), as well as the European Union’s Horizon Europe research and innovation programme under Grant Agreement No.~101070700 (MIRAQLS), and the U.S.~Air Force Research Office under Grant No.~W911NF-22-1-0277. The work conducted at UBC was supported by the Stewart Blusson Quantum Matter Institute (SBQMI) and the Canada First Research Excellence Fund, Quantum Materials and Future Technologies Program. J.S. acknowledges financial support from NSERC (Discovery Grant Program and Quantum Alliance Consortium ``Consortium on Quantum Simulation with Spin Qubits'') and from Defence Canada (Innovation for Defence Excellence and Security, IDEaS). J.S. also acknowledges financial support from the Canada Foundation for Innovation (CFI) John~R.~Evans Leaders Fund and the CFI Innovation Fund. M.E. acknowledges support from the NSERC CREATE in Quantum Computing Program (Grant No.~543245).

\vspace{2em}

\section*{\textbf{METHODS}}

{\textit{Surface cleaning and epitaxial growth}}.
Before the overgrowth, the industrial buffers were cleaned in 2~wt\% HF and 25~wt\% HCl while omitting the final deionized (DI) rinse. As a final step, the buffers were \textit{in situ} annealed in hydrogen at $875~^{\circ}\mathrm{C}$ for 30~min. The growth was performed in a thermal reduced-pressure chemical vapor deposition (RPCVD) system using hydrogen as a carrier gas and centrifugally enriched monogermane (12\% $^{70}$GeH$_4$ in H$_2$ with isotopic purity $>99.9\%$) and monosilane (25\% $^{28}$SiH$_4$ in H$_2$ with isotopic purity $>99.99\%$). The total growth pressure was 20~Torr, with partial pressures of $\sim2$--9~Pa for $^{70}$GeH$_4$ and $\sim6$~Pa for $^{28}$SiH$_4$. The growth temperature was optimized at $550~^{\circ}\mathrm{C}$ following tests in the 600--550~$^{\circ}\mathrm{C}$ range to minimize the formation of Ge islands at the start of regrowth and to promote sharp quantum well (QW) interfaces. 

When initiating the homoepitaxy, an initial boost in $^{28}$SiH$_4$ flow for a few seconds was required to further suppress Ge island formation. When the desired first barrier growth time was reached, the precursor flows were cut for approximately 30~s, during which the reactor was purged with a 2700~sccm hydrogen flow. The growth of the $^{70}$Ge well then followed, succeeded by a 30~s hydrogen purge of the reactor. To ensure a sharp QW/second-barrier interface, a second boost in the $^{28}$SiH$_4$ precursor flow was performed to correct the silicon concentration profile. Finally, a $\sim2$~nm capping layer of $^{28}$Si was grown to protect the purified heterostructure with a stable native oxide.

{\textit{Characterization}}.
Thin specimens suitable for cross-sectional transmission electron microscopy (XTEM) were prepared using a focused ion beam (FIB) in a FEI Helios NanoLab~600 operating with a 30~keV Ga ion beam. TEM imaging was performed on a C-FEG JEOL JEM-Fx2000 operating at 200~kV. To establish the isotopic purity of the material and quantify possible contaminants, laser-assisted atom probe tomography (APT) measurements were conducted using an Invizo~6000 instrument with a picosecond laser at a wavelength of 257~nm, pulse energy of 25--45~pJ, and base temperature of 25~K. Surface roughness of the starting buffers and the $^{28}$Si$^{70}$Ge overgrowth was analyzed using a Bruker Icon FastScan atomic force microscope (AFM) in PeakForce quantitative nanomechanics (QNM) tapping mode over 10--20~$\mu$m scan areas. Surface profiles were corrected in Bruker Nanoscope software using a second-order plane fit. Threading dislocation densities in the heterostructures and commercial buffers were estimated using defect delineation experiments based on a Secco etch (1~part HF, 49~wt\%, to 2~parts 0.15~mol/L K$_2$Cr$_2$O$_7$). Pits were etched at room temperature for 20--60~s and analyzed using AFM. Crystallinity of the QW was evaluated by high-resolution X-ray diffraction (XRD) using a Bruker D8~Discover system equipped with a Cu~K$\alpha_1$ source, a triple-bounce Ge(220) analyzer, and a Ge(220) monochromator. The ($\overline{2}\,\overline{2}\,4$) reflection was used for reciprocal space mapping (RSM) analysis.

{\textit{Device processing}}.
The devices feature titanium/gold bilayer bond pads, aluminum ohmic contacts, an Al$_2$O$_3$ gate dielectric deposited on the Si/SiO$_2$ cap, and a titanium/gold bilayer top gate. A silicon nitride field dielectric was selectively deposited underneath the Al$_2$O$_3$, only below the gate bond pad. The QW region was defined by selective-area reactive ion etching (RIE) to remove the quantum well around the Hall bar (HB) boundary. The mesa etch employed an SF$_6$/O$_2$ chemistry and removed approximately 100~nm of material, including the QW. 

The Al$_2$O$_3$ gate dielectric (22~nm thick) was deposited by thermal atomic layer deposition (ALD) at $280~^{\circ}\mathrm{C}$. The silicon nitride ($\approx250$~nm thick) was deposited by pulsed DC reactive sputtering at room temperature to enhance the electrical robustness of the gate bond pad for wire bonding. Aluminum ohmic contacts were annealed \textit{in situ} during the deposition of the Al$_2$O$_3$ gate dielectric. A schematic cross-section of the device is shown in Fig.~3(b).

{\textit{Transport measurements}}.
Hall bar measurements were performed in a cryo-free variable temperature insert (VTI) at approximately 1.4~K, in the bore of an 8~T superconducting solenoid. The current $I$ between the outer contacts was measured as a function of $V_G$ by applying an oscillating voltage $V_{\mathrm{bias}} = 50~\mu\mathrm{V}$ to one contact and recording the resulting current at the opposite contact using a Stanford SR860 lock-in amplifier.

For magneto-transport measurements, longitudinal ($R_{xx}=V_{xx}/I$) and transverse ($R_{xy}=V_{xy}/I$) resistances were measured by applying an oscillating bias voltage $V_{\mathrm{bias}} = 40~\mathrm{mV}$ across a series resistor $R_s = 4~\mathrm{M}\Omega \gg R_{xx}$, giving $I = V_{\mathrm{bias}}/R_s = 10~\mathrm{nA}$. Voltages $V_{xx}$ and $V_{xy}$ were measured differentially using SR860 lock-in amplifiers. The gate voltage $V_G$ and temperature were held constant while sweeping the perpendicular magnetic field $B_{\perp}$. The gate voltage was swept from 0 to $V_{\mathrm{min}}$, then from $V_{\mathrm{min}}$ back to 0, for progressively more negative $V_{\mathrm{min}}$ values to probe gate-induced charging of interface traps. When necessary, devices were reset by cycling the VTI temperature to 300~K to release trapped charges that shifted the turn-on voltage.

\bibliography{references}

\end{document}